\documentclass[aps,prb,twocolumn,groupedaddress,showpacs,showkeys]{revtex4}

\usepackage{amsfonts}
\usepackage{graphicx}

\def \figwidth{8cm}
\def \figwidthb{4cm}

\begin{document}

\title{Magnetic relaxation of type II superconductors \\ in a mixed state of entrapped and shielded flux}
\author{D. Zola}\thanks{Corresponding author}
\email[\newline e-mail:]{zoldan@sa.infn.it}
\thanks{\newline FAX: +3908965390 \\}
\affiliation{SUPERMAT, INFM Regional Laboratory and Department of
Physics \lq\lq E. R. Caianiello\rq\rq , University of Salerno, via
S. Allende, I-84081 Baronissi, Italy. }
\author{M. Polichetti}
\affiliation{SUPERMAT, INFM Regional Laboratory and Department of
Physics \lq\lq E. R. Caianiello\rq\rq , University of Salerno, via
S. Allende, I-84081 Baronissi, Italy. }
\author{C. Senatore}
 \affiliation{SUPERMAT, INFM Regional Laboratory and Department of Physics \lq\lq E. R. Caianiello\rq\rq
, University of Salerno, via S. Allende, I-84081 Baronissi, Italy.
}
\author{S. Pace}
\affiliation{SUPERMAT, INFM Regional Laboratory and Department of
Physics \lq\lq E. R. Caianiello\rq\rq ,  University of Salerno,
via S. Allende, I-84081 Baronissi, Italy. }

\date{\today}

\begin{abstract}
The magnetic relaxation has been investigated in type II
superconductors when the initial magnetic state is realized with
entrapped and shielded flux (ESF) contemporarily. This flux state
is produced by an inversion in the magnetic field ramp rate due to
for example a magnetic field overshoot. The investigation has been
faced both numerically and by measuring the magnetic relaxation in
BSCCO tapes. Numerical computations have been performed in the
case of an infinite thick strip and of an infinite slab, showing a
quickly relaxing magnetization in the first seconds. As verified
experimentally, the effects of the overshoot cannot be neglected
simply by cutting the first 10-100 seconds in the magnetic
relaxation. On the other hand, at very long times, the  magnetic
states relax toward those corresponding to field profiles with
only shielded flux or only entrapped flux, depending on the
amplitude of the field change with respect to the full penetration
field of the considered superconducting samples. In addition, we
have performed numerical simulations in order to reproduce the
relaxation curves measured on the BSCCO(2223) tapes; this allowed
us to interpret correctly also the first seconds of the $M(t)$
curves.
\end{abstract}

\pacs{74.25.Ha, 74.25.Qt, 74.72.-h} \keywords {High T$_c$
Superconductors; Magnetic relaxation, Flux creep.}

\maketitle


\section{\label{Introduction}Introduction}
In type II superconductors, at temperatures $T \neq 0$, the
magnetization relaxes approximately logarithmically on time ($t$)
because of the thermally activated motion of vortices (flux
creep). This behaviour can be understood, at first sight, within
the Anderson Kim model (AKM)
\cite{PR131(1963)2486,PRL9(1962)309,RMP36(1964)39,PR181(1969)682}.
In conventional superconductor the experimental results are well
reproduced in the framework of the AKM, whereas in high
temperature superconductors (HTS), deviations from the logarithmic
decay are observed, especially in Bi-based materials
\cite{PRB40(1989)7380,PRB40(1989)10882,PC249(1995)144,PC332(2000)374}.
Several models have been proposed in order to explain the
non-logarithmic relaxation
\cite{RMP66(1994)1125,PRL62(1989)1415,PRL63(1989)2303,Cryo30(1990)563,RMP68(1996)911}.
The theory of collective creep, extensively reviewed by Blatter et
al. \cite{RMP66(1994)1125}, predicts that the current density
($J$) relaxes according to the so called $\lq\lq$interpolation
formula". As in the case of the Bean fully penetrated critical
state, the magnetization can be assumed proportional to the
persistent current, leading to:
\begin{equation}\label{eq1}
 M(t,T) = \frac{M_i}{\left[1+\frac{\mu k_B T}{U_0}\ln\left(\frac{t}{t_0}\right)\right]^{1/\mu}}
\end{equation}
where $M_i$ is the initial value of the magnetization, $k_B$ is
the Boltzmann constant and $U_0$ is the pinning activation energy.
The exponent $\mu$ is a parameter and its value depends on the
different creep regimes; $t_0$ is a characteristic time depending
on temperature, magnetic field, sample geometry and the fluxon
attempt frequency for jumping out the pinning centres. By defining
the normalized creep rate ($S$):
\begin{equation}\label{eq2}
S=\frac{1}{M}\frac{dM}{d \ln t}
\end{equation}
the equation (\ref{eq1}) immediately leads to:
\begin{equation}\label{eq3}
S=\frac{- k_B T}{\left[ U_0 + \mu k_B T \ln \left(t/t_0
\right)\right]}
\end{equation}
The equation (\ref{eq3}) is employed to evaluate experimentally
the pinning activation energy and the exponent $\mu$. For this
reason, magnetic relaxation measurements are extensively used to
investigate the flux creep in superconductors
(for a review, see \onlinecite{RMP68(1996)911} and references therein).\\
\indent Usually, in magnetic relaxation measurements ($M(t)$), an
external magnetic field $H_a$ ramps up to a fixed value $H_0$ with
finite sweep rate $dH_a/dt$, then the magnetization is measured as
function of time (typically for about $10^3$ seconds) keeping the
external field at the fixed value. Ramping the external field up
to $H_0$, that is chosen higher than the full penetration field
$H_p$ of the superconductor, screening persistent currents
(clockwise with respect to the external field versus) flow
everywhere in the superconductor. If the magnetic field is firstly
increased and then slightly reduced, both clockwise and
counterclockwise persistent currents flow in the sample. In this
case, the measured magnetization results from a region with
entrapped flux close to the surface and a region with shielded
flux in the inner part of
the superconductor (ESF state).\\
\indent This complicated state can be easily generated when the
external field ramp is stopped and a magnetic field overshoot
occurs. This means that, at the nominal stop of the external field
ramp, the field exceeds the target value $H_0$, reaching it
usually after few seconds. This overshoot can produce an entrapped
flux zone close to the surface, which can appreciably affect the
relaxation process. In particular, Jirsa et al.
\cite{PC207(1993)85,JLTP78(1990)179} showed that, for a
superconducting slab of thickness $10^{-4}$ m in a parallel field
$H_0$ = 0.5~T, an overshoot of only 1.5~mT leads to an initial
magnetization $M_i^{ov}$, whose value is about one third of the
one computed in the absence of the overshoot. However, the
depressed magnetization $M_{ov}(t)$ relaxes with time converging
to the ideal $M_{id}(t)$ curve computed in absence of overshoot.
Therefore, the initial value of the magnetization, occurring in
the absence of the overshoot, is determined approximately by
extrapolating it from the long time $M(t)$ curve.\\
\indent However, starting from the ESF state, the field profile
evolution that leads to the joint of the two curves is still
unclear. On the other hand, it is not possible to determine
experimentally when the $M(t)$ curve approaches to the ideal
relaxation and, thus, it is usually adopted the experimental
procedure of cutting the first
10-100 seconds in the experimental $M(t)$. \\
\indent In order to justify this experimental procedure, we can
consider a slab of thickness $2\,d$ and critical current density
$J_c$ analyzed in the framework of the Bean model. If an overshoot
occurs after the application of an external field higher than the
full penetration field ($H_p = J_c d$), the magnetization of the
slab in the framework of the Bean model, is:
\begin{eqnarray}
M &=& M_{en} + M_{sh} \\
M_{en} &=& (1/4)(H_{ov}^2/H_p) \\
M_{sh} &=&-(1/2)((H_p^2-H_{ov}^2)/H_p)
\end{eqnarray}
where $M_{en}$ is the magnetization due to the entrapped flux,
$M_{sh}$ is the magnetization due to the shielded flux and
$H_{ov}$ is the amplitude of field overshoot. If $H_{ov} \ll H_p$,
the magnetization due to the entrapped flux is small and thus it
can be considered negligible after a long enough time. In a low
T$_c$ superconducting slab, with $d = 0.1$~mm and $J_c =
10^{10}$A/m$^2$,  the full penetration field is $H_p =$ 0.63~T and
the usual characteristic time $t_0$ is about 10 seconds.
Therefore, for a few mT overshoot, it is commonly believed that
the experimental $M(t)$ measured 100 seconds after the nominal
stop of the external magnetic field resembles the relaxation from
a fully shielded state (or a fully entrapped state). Nevertheless,
depending on the temperature and the applied magnetic field, $H_p$
can become comparable with $H_{ov}$,
drastically affecting also the long-time magnetic relaxation.\\
\indent To extend the relaxation analysis to the time window
affected by the overshoot, Jirsa et
al.\cite{PC207(1993)85,JLTP78(1990)179} have shown that it is
possible to use magnetic hysteresis loop data measured at
different field sweep rates. They have shown how the magnetization
measured at different sweep rates can be converted into magnetic
relaxation data, substantially extending the
time window to the short times, typically down to  10$^{-2}$ s.\\
\indent Other complications in the analysis of relaxation
measurements can also arise from the sample geometry and the
anisotropic properties of the material. In fact, in HTS samples,
magnetic relaxations are usually measured with the field
orientation perpendicular to the largest face of the sample. In
this geometry, the demagnetization effects could be neglected only
for measurements performed at fields much higher than $H_p$. Since
an overshoot changes the direction of the current and the magnetic
field value on the edge of a flat superconductor, geometry effects
are supposed to be
significatively altered in the magnetic relaxation measurement.\\
\indent In this work we have investigated the magnetic relaxation
starting
from a state with entrapped and shielded flux.\\
\indent In the next section, we will discuss the
integro-differential equation employed in the numerical
computation of the $M(t)$ curves. In the Section \ref{sec3}, we
show the numerical simulations of the magnetic relaxation and the
time evolution of the field profiles for samples in shape of slab
and thick
strip.\\
\indent The magnetic relaxations in BSCCO(2223) have been
experimentally investigated when the effects of a magnetic field
overshoot in the $M(t)$ are not negligible. Finally, in the
Section \ref{sec4}, the experimental measurements are analyzed and
compared with the numerically computed results.

\section{\label{sec2}Numerical computations}
In order to analyze the magnetic relaxation of a superconductor in
an external magnetic field $H_0$, we numerically solved an
integro-differential equation for the current density $J$ in a
slab in parallel field and in a thick strip in perpendicular
field\cite{PRB54(1996)4246}. As developed by Brandt in a series of
works
\cite{PRB54(1996)4246,JAP84(1998)5652,PRB58(1998)6506,PRB58(1998)6523,PRB64(2001)214506},
in a long strip of width 2$a$ (along $y$ axis) and thickness 2$d$
(along $z$ axis) placed into a homogeneous magnetic field,
perpendicular to the largest face of the strip, the applied field
induces surface and bulk currents. The current flows along the
sample length (i.e. $x$ axis) due to the symmetry of the strip.
The induced current density $\mathbf{J}= J(y,z) \mathbf{i}$
generates a magnetic field $\mathbf{H}$ which has $y$ and $z$
components. In this model it is assumed that
$\mathbf{B}=\mu_0\mathbf{H}$ and thus, $H_{c1}$ and the reversible
magnetization ($M_{rev}$) are neglected. Since $\mathbf{B}=
 \mathbf{\nabla}\times\mathbf{A}$, where
 $\mathbf{A}$ is the vector potential, it is possible
 to write for this geometry
a 2D Poisson equation in the Coulomb gauge
\begin{equation}\label{eq2:1}
  \mu_0J =-\nabla^2A
\end{equation}
The current density flows only in the strip and thus the vector
potential could be written as a sum of two terms $A = A_a + A_J$,
where $A_a$ is the vector potential related to the applied
magnetic field,  ($A_a = \left[\mathbf{r}\times
\mathbf{B}\right]_x
  =yB_a$), and $A_J$ is related to the current induced in the strip.
Since $B_a$ is constant in the specimen, the general solution of
the (\ref{eq2:1}) is:
\begin{equation}\label{eq2:3}
  A(\mathbf{r})= -\mu_0\int_Sd^2r'Q(\mathbf{r},\mathbf{r}')
  J(\mathbf{r}',t)-yB_a
\end{equation}
where $\mathbf{r}= (x,y)$,  $\mathbf{r}'= (x',y')$ and
$Q(\mathbf{r},\mathbf{r}')$ is the integral kernel defined as:
\begin{equation}\label{eq2:4}
Q(\mathbf{r},\mathbf{r}')=\frac{1}{2 \pi}\log
\left|\frac{\mathbf{r}-\mathbf{r}'}{r_0}\right|,
\end{equation}
in which $r_0$ is an arbitrary constant length that can be chosen
equal to 1. The integration is performed on the cross section of
the strip $S$. The current density, is  obtained formally from
\cite{PRB54(1996)4246}:
\begin{equation}\label{eq2:5}
  J(\mathbf{r},t)=-\frac{1}{\mu_0}\int_{S'}d^2r'Q^{-1}
  (\mathbf{r},\mathbf{r}')\left[A(\mathbf{r}',t)-y\dot{B}_a\right].
\end{equation}
 Here $Q^{-1}(\mathbf{r},\mathbf{r}')$ is the inverse
  kernel defined by:
 \begin{equation}\label{eq2:6}
  \int_{S}d^2r'Q^{-1}(\mathbf{r},\mathbf{r}')
  Q(\mathbf{r}',\mathbf{r}'')=\delta(\mathbf{r}-\mathbf{r}'').
\end{equation}
By using the relation $ \mathbf{E} = -\mathbf{\nabla}_x\phi -
\dot{\mathbf{A}}$ where $\phi$ is the scalar potential, we obtain
\begin{equation}\label{eq2:7}
  \dot{J}(\mathbf{r},t)= \frac{1}{\mu_0}\int_{S'}d^2r'Q^{-1}
  (\mathbf{r},\mathbf{r}')\left[E(J)-y'B_a(t)\right]
  \end{equation}
In the limit $d \gg a$ (slab geometry), the previous equation
becomes an one-dimensional equation:
\begin{equation}\label{eq2:7a}
  \dot{J}(\mathbf{r},t)= \frac{1}{\mu_0}\int_{0}^{a} dy'Q^{-1}_{slab}
  (y,y')\left[E(J)-y'B_a(t)\right]
  \end{equation}
Taking into account the symmetry of the strip and slab geometries,
the kernel in the case of the strips is given by:
\begin{equation}
Q_{strip}  =  \frac{1}{4 \pi} \ln \frac{\left(y_{-}^{2}+
z_{-}^{2}\right)\left(y_{-}^{2}+
z_{+}^{2}\right)}{\left(y_{+}^{2}+
z_{-}^{2}\right)\left(y_{+}^{2}+ z_{+}^{2}\right)}
\end{equation}
where $y_{\pm} = y \pm y'$ and $z_{\pm} = z \pm z'$. For the slab
it results
\begin{equation}
Q_{slab}  =  \frac{1}{2} (|y-y'|-|y+y'|)=-\min(y,y')
\end{equation}
In our simulations, we do not consider a transport current but
only an external magnetic field and for this reason the term
$\mathbf{\nabla}_x\phi$ has been dropped out. To solve the
integral equation for $\dot{J}$ we choose the widely used relation
\cite{APL56(1990)1700}:
\begin{equation}\label{eq2:8}
  E = E_c\left(\frac{J}{J_c}\right)^n
\end{equation}
where $J_c$ is the critical current density. However, the Brandt
method can be used with different
$E-J$ relationship \cite{PRB54(1996)4246}.\\
\indent The current density profiles in the strip has been
obtained by integrating the equation (\ref{eq2:7}), whereas for
the slab the equation (\ref{eq2:7a}) has been solved. For the
strip, the functions $J$, $E$ have been tabulated on a 2D grid
with equidistant points $y_k = (k-1/2)a/N_y \quad (k = -N_y+1,
\cdots, 0, \cdots, N_y)$  and $z_l = (l -1/2)d/N_z \quad (l=
-N_z+1,\cdots,0,\cdots,N_z$), where $N_z = d/aN_y$ is chosen.
Labelling the points $(y_k,z_l)$ by an index $i$, with $i=1,2,
\cdots, N$ and $N = N_yN_z$, the function $J(y,z,t)$ becomes the
time dependent vector $J_i(t)$  with $N$ coordinates and
$E(y,z,t)= E_c(J/J_c)^n$ becomes a vector with $N$ coordinates.
Moreover, the integral kernel $Q(y,z,y',z')$
becomes an $N \times N $ matrix $Q_{i,j}$.\\
\indent The numerical form of the equation (\ref{eq2:5}) is
\begin{eqnarray}\label{eq2:9}
  J_i(t+\Delta t) &=& J_i(t) + \frac{\Delta t}{\mu_0 \Delta y \Delta
  z}\sum_j^N Q^{-1}_{i,j} \left[E_j(t)-y\dot{B}_z\right] \nonumber \\
  \qquad & \qquad & \textrm{for}\quad
  i=1,{} \cdots {},N
\end{eqnarray}
where $\Delta y =a/N_y$ and $\Delta z = d/N_z$ are respectively
the steps in the 2D grid used to tabulate the cross section of the
thick strip. The numerical integration of the 1D equation for a
slab follows similar rules.\\
\indent The time integration of this system of non-linear
differential equations for $J_i(t)$ has to follow some
prescriptions. First of all, the integration starts with the
initial condition $J_i(0)=0 $; in addition, the time step $\Delta
t$ is chosen inversely proportional to the maximum value of the
resistivity $ \rho_i = E_i/J_i$. Brandt\cite{PRB54(1996)4246} uses
the following relation in his computations: $\Delta t =
c_1/[\max(\rho_i(t))+c_2]$ with $c_1 = 0.3/(N_y^2n)$, $n$ is the
exponent in the $E-J$ law and $c_2 = 0.01$. In our computations we
do not use normalized quantity and we have observed that this
choice depends on the value of $J_c$ and the time derivative of
the external magnetic field. In our computations we used different
values for $c_1$ and $c_2$ in order to make stable the numerical
algorithm: $c_1 = 0.003/[(N_x^2n)\sqrt{\overline{\dot{B_a}^2}}$
where $\overline{\dot{B_a}^2}$ is the temporal mean value of
$\dot{B}_a$ and $c_2 = 1$. \\
\indent Finally, $Q_{i,j}=\ln|\mathbf{r}_i-\mathbf{r}_j|$ has a
logarithm divergence when $\mathbf{r}_i$ approaches
$\mathbf{r}_j$. In order to avoid this singularity for $i=j$ the
expression for the kernel is changed with
$(1/2)\ln[(\mathbf{r}_i-\mathbf{r}_j)^2+\epsilon^2]$ where
\cite{PRB58(1998)6506}:
\begin{eqnarray*}
\epsilon^2 & = & \exp[\ln(\Delta y^2 + \Delta z^2)-\ln(4)-3 + \\
& + & \frac{\Delta y}{\Delta z} \arctan(\frac{\Delta z}{\Delta
y})+
(\frac{\Delta z}{\Delta y})\arctan(\frac{\Delta y}{\Delta z})].\\
\end{eqnarray*}
In our computations, the magnetization is calculated by
\begin{eqnarray}
M &=&  4 \int_0^a dy\int_0^d dz
J(y,z)y \qquad \textrm{for a strip}\\
M & =& 2 \int^{a}_{0} dy J(y)y  \qquad \qquad \qquad \textrm{for a
slab}
\end{eqnarray}
Since the magnetic relaxation is simulated on $10^5\div 10^6$
seconds, we reduce the number of computed points calculating the
$(t_i,M_i)$ data accordingly to the relation:
\begin{eqnarray}
 t_i &=& t_{i-1}+ \exp(\log(t_R)/N_R)\\
 M_i &=& M(t_i)
\end{eqnarray}
where $t_R$ is the total time of the computed relaxation and $N_R$
is the total number of the computed data.

\section{\label{sec3}Numerical results}
\begin{figure}[tbp]
\includegraphics[width=\figwidth, clip]{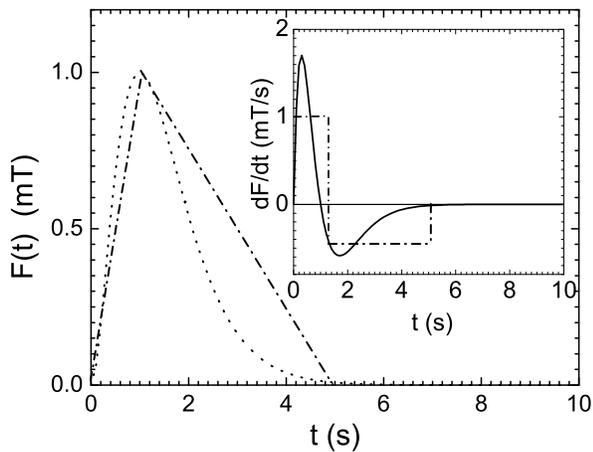}
\caption{Time dependence of the external magnetic field during the
field overshoot. The time origin corresponds to the nominal field
stop. A triangular overshoot (dashed-dotted line) and an overshoot
given by the function $ \Delta H_{ov} = H_{ov} (t/t_{ovm})^2 \exp
(2(1-t/t_{ovm}))$ (dotted line) are shown. In the inset, the time
derivatives of the two overshoot functions are shown.} \label{fov}
\end{figure}
In this section we discuss the numerical results obtained for the
slab and the thick strip. In our computations we have used a strip
with aspect ratio $(a/d)$ equal to 10 and $2a = 10^{-3}$~m and a
slab with $2a = 10^{-4}$~m, with the critical current density
($J_c$) ranging from 10$^6$ A/m$^2$ to 10$^{9}$ A/$m^2$. The
current-voltage characteristic is the usual power law given by $E
= E_c(J/J_c)^n$, where $E_c = 10^{-4}$~V/m and the employed
exponent $n$ is chosen equal to 15 for the large creep case and $n
= 105$ in the Bean limit case.\\
\indent In order to study the relaxation from a ESF state,
different magnetic field ramps have been taken in account. For
each ramp, the external magnetic field $H_a$ increases linearly on
time, with a sweep rate ($\dot{H}_a$) of 1~mT/s, up to a nominal
fixed value $H_0$. The time when $H_a$ has nominally reached
$H_{0}$ is taken as time origin of the magnetic relaxation. As the
external magnetic field reaches $H_{0}$ different situations are
taken into account:
\begin{itemize}
\item [a)] $H_a$ is stopped immediately (ideal case); \item [b)]
$H_a$ has a triangle overshoot (triangle overshoot); \item [c)]
$H_a$ has an overshoot with a smoothed field stop (exponential
overshoot).
\end{itemize}
In the case b), the magnetic field increases in  $t_{ovm}$ seconds
by an amplitude $H_{ov}$, then it decreases by the same quantity
in the subsequent $t_{ov}$ seconds (triangle overshoot). After
this, the external field is immediately stopped and the magnetic
relaxation starts. In the case c), the overshoot has been
simulated {by means of the function $ F_{ov}(t) = H_{ov}
(t/t_{ovm})^c \exp (c(1-t/t_{ovm}))$; for $t=t_{ovm}$ the
overshoot reaches the maximum value. The two different functions
employed to simulate an overshoot are shown in Fig. \ref{fov} a).
For the triangular overshoot, we set $H_{ov} = 1$ mT, $t_{ovm} =
1$ s, $t_{ov} = 5$s. In the case of exponential overshoot, we used
$H_{ov} = 1$ mT, $t_{ovm} = 1$ s, $c = 2$. In the inset of the
same figure, the time derivative of the overshoot functions are
plotted, since the field ramp
derivative is actually used in the integration of the diffusion equation.\\
\begin{figure}[tbp]
\includegraphics[width=\figwidth, clip]{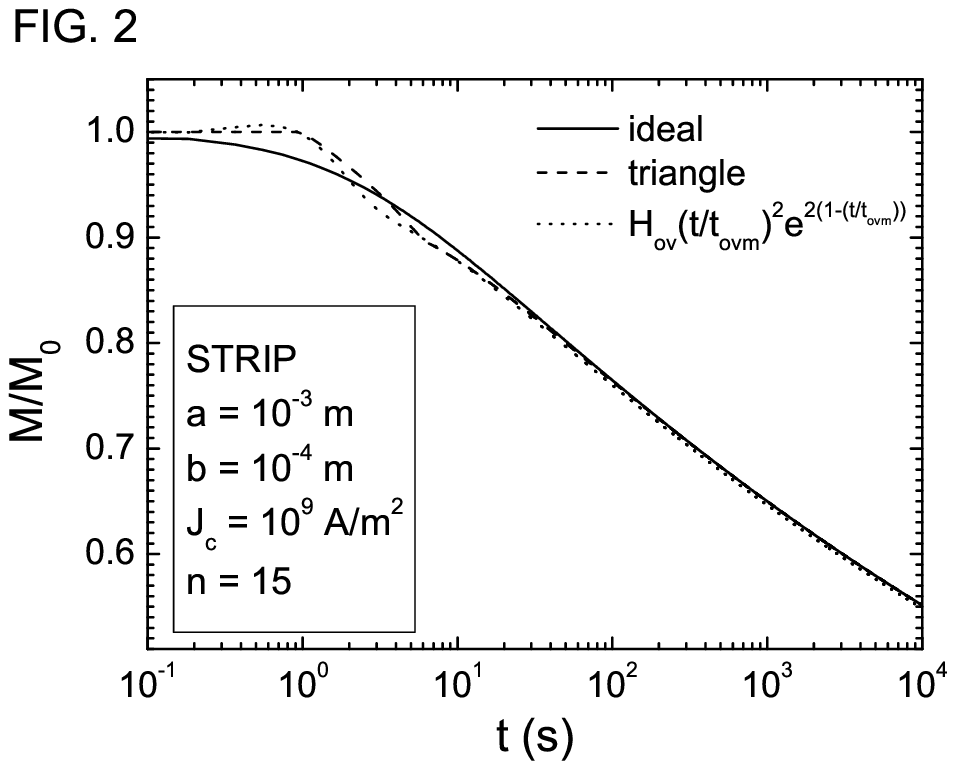}
\caption{Magnetic relaxation curves computed for different
magnetic field ramps in a thick strip.} \label{calcmtjc9}
\end{figure}
\indent We  have initially computed the magnetic relaxations for a
strip in perpendicular magnetic field (perpendicular geometry) by
simulating a case analogous to the one discussed in the work of
Jirsa et al \cite{PC207(1993)85}. In our computation  $J_c =
10^9$~A/m$^2$ and the critical exponent is $n = 15$. We are
considering a superconductor with large critical current density
but with large creep. The external field ramps with a sweep rate
of 1~mT/s up to 0.2~T, which is a value well above the full
penetration field of the strip. Indeed, looking at the field
profile we have verified that the strip is  fully penetrated for
fields higher than 0.10~T. As shown in Fig. \ref{calcmtjc9}, also
if the overshoot does not occur in the field ramp, the
magnetization decays non-logarithmically, especially at short time
($\le$ 10~s). This result is expected due to the power law in the
$E-J$ relationship which involves a logarithmic dependence of the
pinning energy on the current density. In the same figure, a
magnetic relaxation curve is shown as computed for a field ramp
which has a triangular overshoot. In this case, the external
magnetic field ramps up to 0.2~T. After this, the field overshoot
occurs with an amplitude of $H_{ov}=1$~mT. The field overshoot
reaches its maximum  one second after the external field should
have been stopped at the nominal target value. The external field
goes down to the nominal value of 0.2~T after 5 seconds. Also in
this case, when the magnetic field is stopped at the fixed value
of 0.2~T the time derivative of $H_a$ is instantaneously zero. A
more realistic situation have been considered by computing the
magnetic relaxation for the case c) where $H_{ov} = 1$~mT,
$t_{ovm} = 1$~s and $c = 2$. The case c) is effectively realized
in experiments, where the field cannot be
stopped instantaneously and the overshoot shape is rounded. \\
\indent As shown in Fig. \ref{calcmtjc9}, the magnetization curves
in the  cases b) and c)have an initial values $M_i$ larger than in
the ideal case. In fact, when the overshoot occurs, the
magnetization does not relax during the first seconds, since the
magnetic field continues to increase. The largest value of the
initial magnetization is obtained in the case of an exponential
overshoot; indeed, the electrical field induced in the
superconductor in the first seconds is larger than in the other
cases (see also Fig. \ref{fov}). When the external magnetic field
rate reverses, the magnetization quickly decreases, because of the
flux coming out from the surface, and after 5~seconds $M$ has lost
the 12\% of the initial value. The decay during the first 5
seconds depends on the shape of the field overshoot as function of
the time. In the triangular case, the magnetization curve shows a
convex concavity, whereas in the case c) the curvature is concave.
After 5 seconds, the external field is practically constant and
the magnetic relaxation effectively starts; for $t$ larger than
100 seconds the three curves join together. These computations
confirm also in perpendicular geometry,  the results found for
parallel geometry in Ref.\onlinecite{PC207(1993)85}. However in
this case the field overshoot amplitude is 1\% of the full
penetration field. In the next section we will consider situations
where the induced ESF state strongly affects the magnetic
relaxation.

\subsection{ESF state in slab}
\begin{figure}[htb]
\includegraphics[width=\figwidth, trim = 0cm 0cm 0cm 0cm, clip]{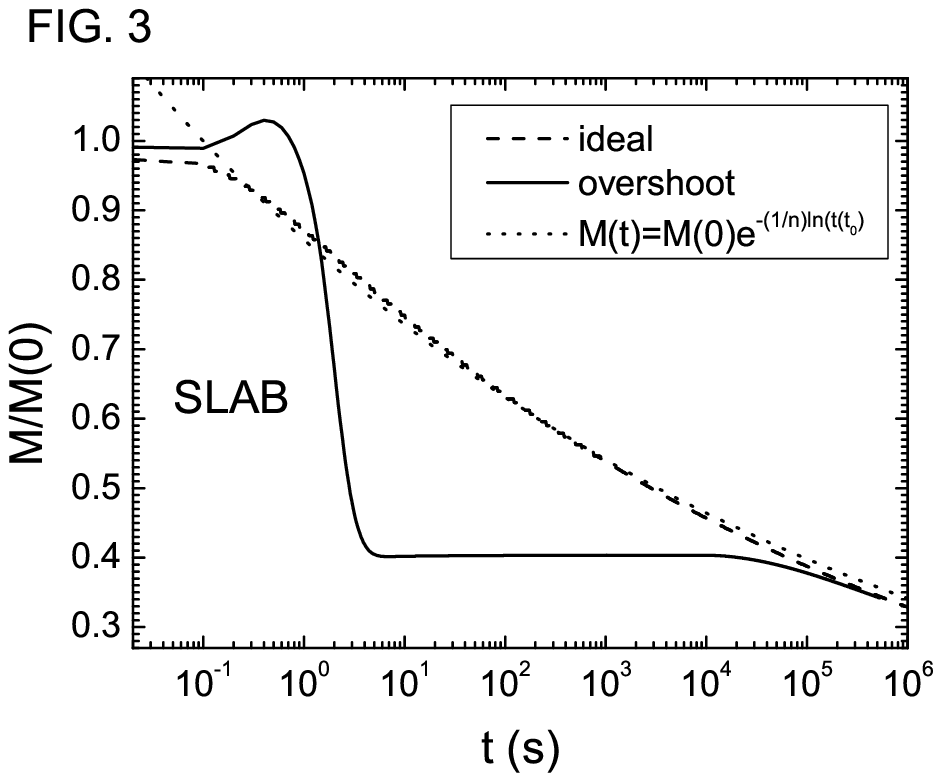}
\caption{Magnetic relaxation curves computed for different
magnetic field ramps in a slab. The dotted line is the relaxation
given by the analytical formula reported in the text.}
\label{calcmtnslab}
\end{figure}
\begin{figure}
\includegraphics[width=\figwidth, trim = 0cm 0cm 0cm 0cm, clip]{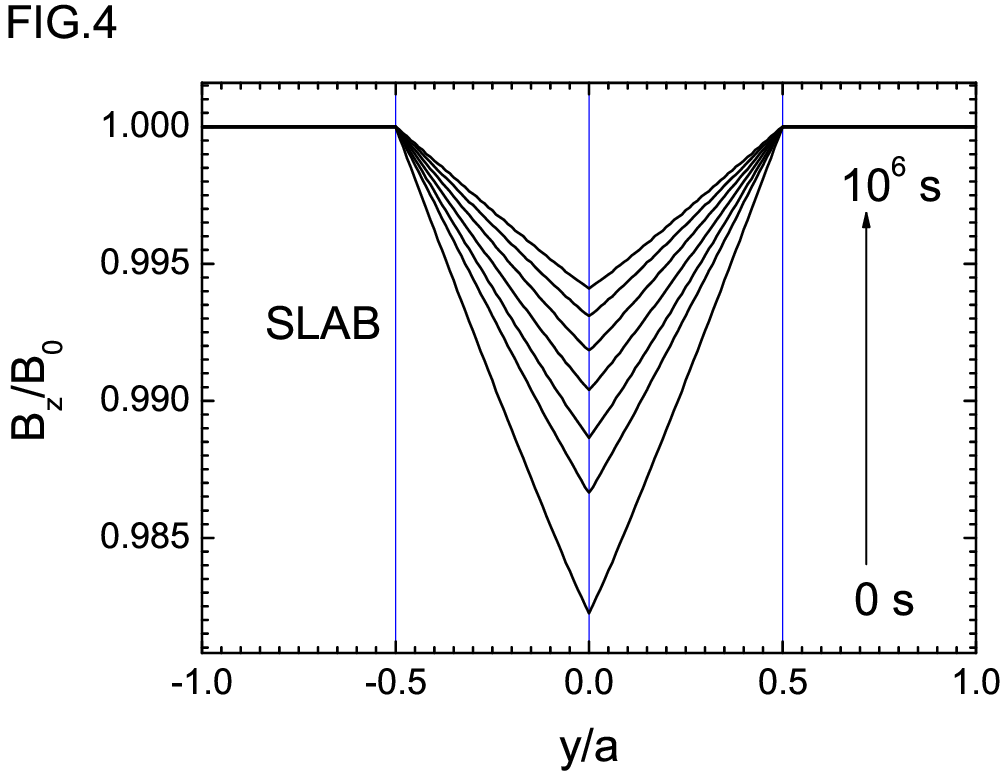}
\caption{Time evolution of the magnetic field profiles in a slab;
the initial field profile is achieved without field
overshoot.}\label{profilislabidn}
\end{figure}
\begin{figure}[ptb]
\includegraphics[width=\figwidth,  clip]{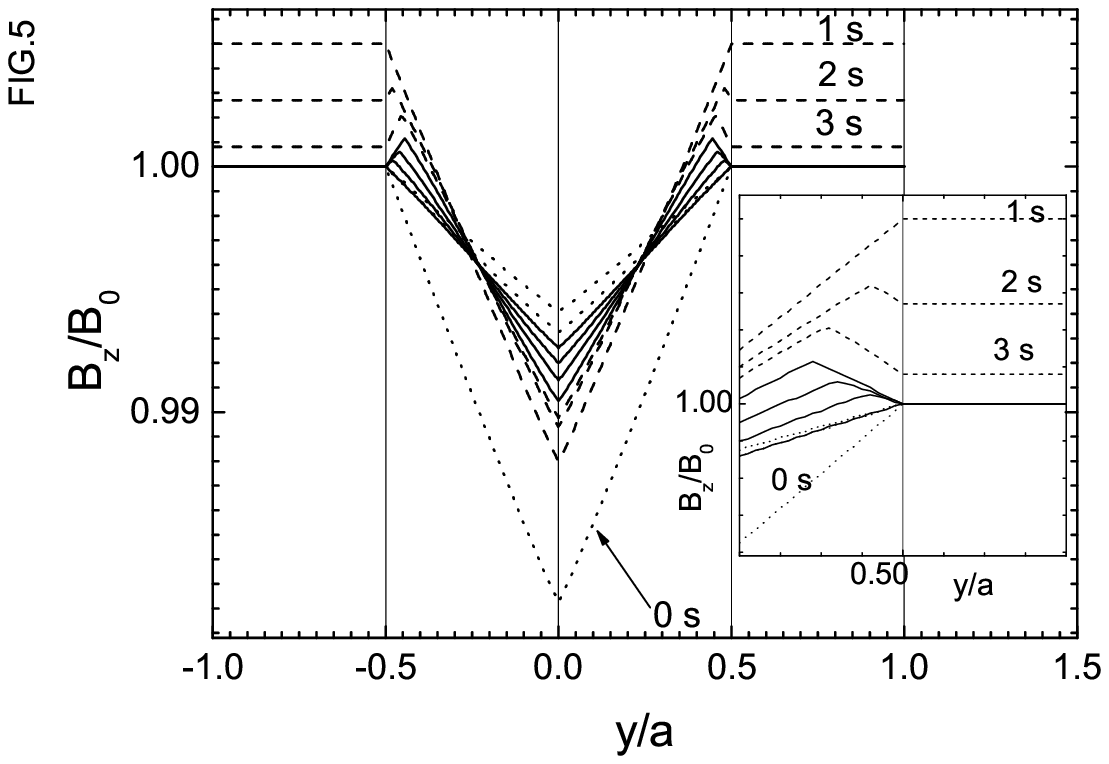}
\caption{Time evolution of the magnetic field profiles in a slab,
when a field overshoot occurs. The dotted lines represent the flux
profiles which fully resemble the ideal profiles. The dashed lines
represent the profile during field ramp rate reversing. Continuous
lines show the field profiles in the time windows where the
magnetization is nearly constant. In the inset a detail of the
profile close to the slab surface is shown.}\label{profilislabovn}
\end{figure}
\indent Here, we discuss the magnetic relaxation starting from an
ESF state in the case of a slab in parallel field. In Fig.
\ref{calcmtnslab}, two computed $M(t)$ curves are shown; the
initial magnetic state is obtained by ramping the external field
both in the ideal way (without overshoot) and with an overshoot of
1~mT (dashed curve) In the same figure, it is shown the magnetic
relaxation (dotted line) for a superconducting slab, according to
the relation given in Ref. \onlinecite{RMP68(1996)911}, where it
is assumed that the pinning energy depends logarithmically on the
current density:
\begin{equation}\label{eq10}
  M(t)=M(0)\exp\left(-\frac{1}{n}\ln\left(\frac{t}{t_0}\right)\right)
\end{equation}
The dimension of the slab used for the computations is $2a =
10^{-4}$m, the critical current density $J_c = 10^8$~A/m$^2$ and
the exponent $n = 15$. In this case the full penetration field of
the slab is $H_p = 6.3$~mT and thus it is of the same order of
magnitude with respect to the  overshoot (1 mT). As shown in Fig.
\ref{calcmtnslab}, the computed ideal curve is approximated quite
well by the analytical relation in the time range from 10 to
10$^4$ seconds, whereas it wanders off each other at very short
and very long times. On the other hand, we observe as the
overshoot has effects on long time up to $10^{5}$~s (dashed
curve). In the first 5 seconds, the magnetization looses 60\% of
the initial value due to the inversion of the flux profile close
to the slab surface. In the subsequent 10$^4$ seconds the
magnetization practically does not relax, and after this time the
relaxation rate increases. After 10$^6$ seconds the magnetization
computed with an
ideal ramp and the curve computed with a field overshoot take the same value.\\
\indent At this point, it is necessary to investigate if  the
magnetization computed for time larger than $3.0 \ 10^5$~s in both
the cases, corresponds to the same magnetic state. In order to
answer this question, we have computed the magnetic field profiles
as a function of time. In Fig. \ref{profilislabidn} and in Fig.
\ref{profilislabovn}, the field profiles computed for both the
cases are shown. In particular, in Fig. \ref{profilislabidn}, the
profiles of the relaxation in a slab are shown, reproducing the
usual Bean results. On the other hand, the profiles computed in
the case of a relaxation from an ESF state, obtained by using the
exponential overshoot, (Fig. \ref{profilislabovn} show  that
during the first 5 seconds the profile changes (dashed line) as a
consequence of the field decreasing. The evolution of the profiles
during the first 5 seconds has some difference in comparison with
the classic Bean profile, where $J_c$ is constant and independent
on the applied electrical field. In our case, while the flux is
expelled on the surface, in the inner part of the slab the profile
relaxes. This occurs because of the finite exponent $n$ which
leads to a large creep. On the contrary, for the Bean model, the
field profile, in the inner part of the slab, remains
frozen during the field decreasing.\\
\indent Starting from the fifth second the field profile relaxes
overall in the slab and after $10^5$~seconds the magnetic profile
becomes the ideal one. In Fig. \ref{profilislabovn} the field
profiles which resemble the ideal ones are shown by dotted line.
By means of our numerical simulations we have shown that the same
magnetization value found in the two $M(t)$ curves corresponds to
the same magnetic state. In Fig. \ref{profilislabovn}, we observe
also that the maximum of the field profile, due to the field ramp
rate reversing, moves towards the slab edges during the
relaxation. At the same time, the entrapped magnetization is
reduced down to zero. Therefore the ESF state has
relaxed towards a fully shielded state.\\
\indent Increasing the amplitude of the overshoot, we expect that
 the ideal relaxation and the relaxation from a ESF
state will coincide at longer times. Nevertheless, as the region
with entrapped flux prevails on the shielded region, the flux
profile relaxes towards a fully entrapped state.

\subsection{ESF state in strip}
\begin{figure}
\includegraphics[width=\figwidth, trim = 0cm 0cm 0cm 0cm, clip]{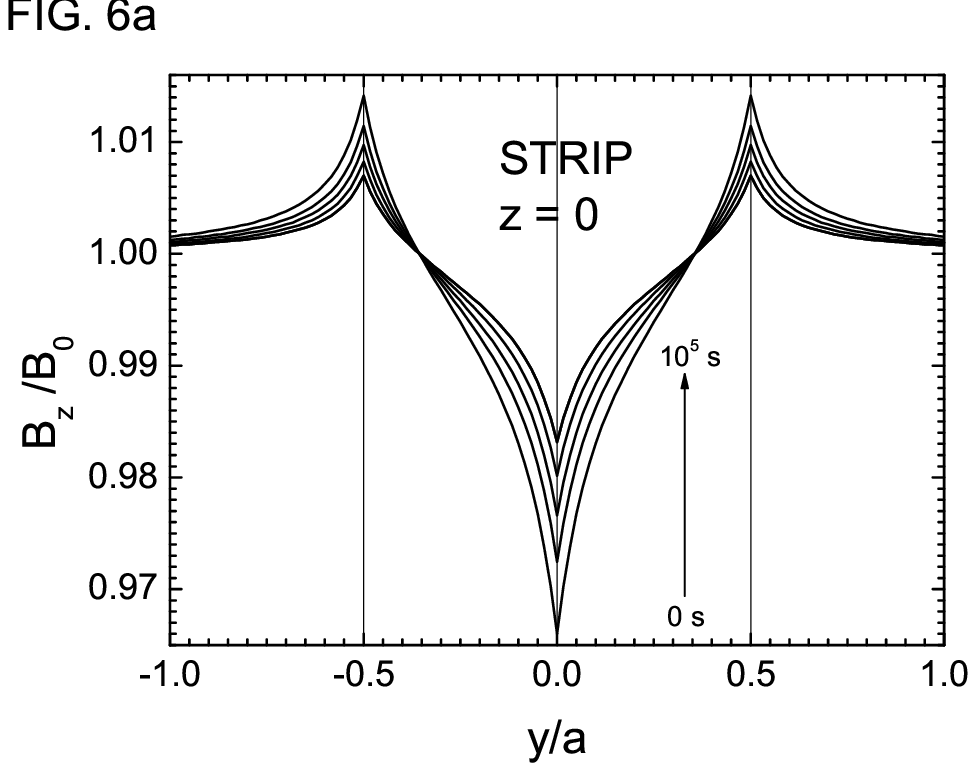}
\includegraphics[width=\figwidth, trim = 0cm 0cm 0cm 0cm, clip]{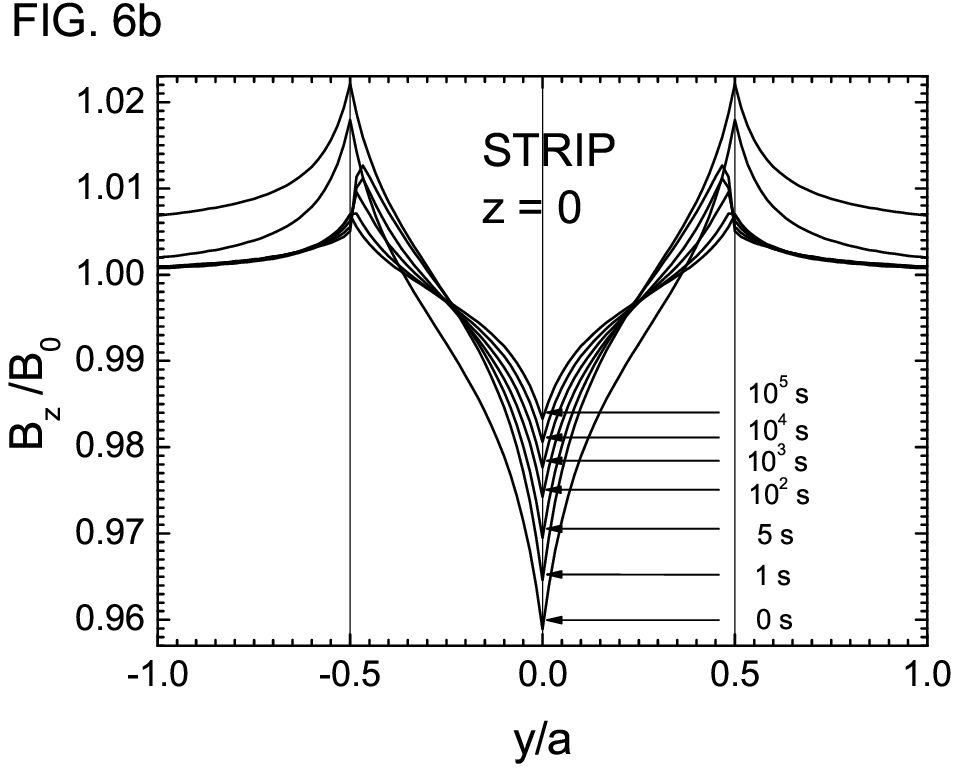}
\caption{Evolution field on time for a thick strip. The field
profiles computed in the ideal case are shown on the upper frame.
The field profiles computed when an overshoot occurs in the field
ramp are shown on the lower frame.}\label{profilistripovz0}
\end{figure}
\begin{figure}
\includegraphics[width=\figwidthb , trim = 0cm 0cm 0cm 0cm, clip]{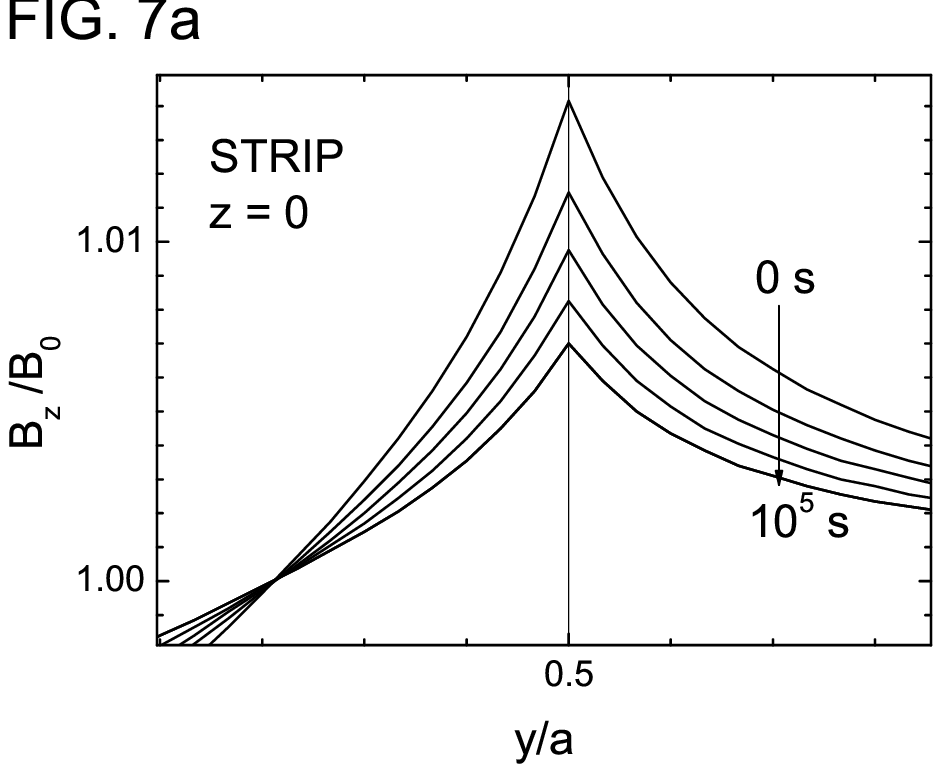}
\includegraphics[width=\figwidthb, trim = 0cm 0cm 0cm 0cm, clip]{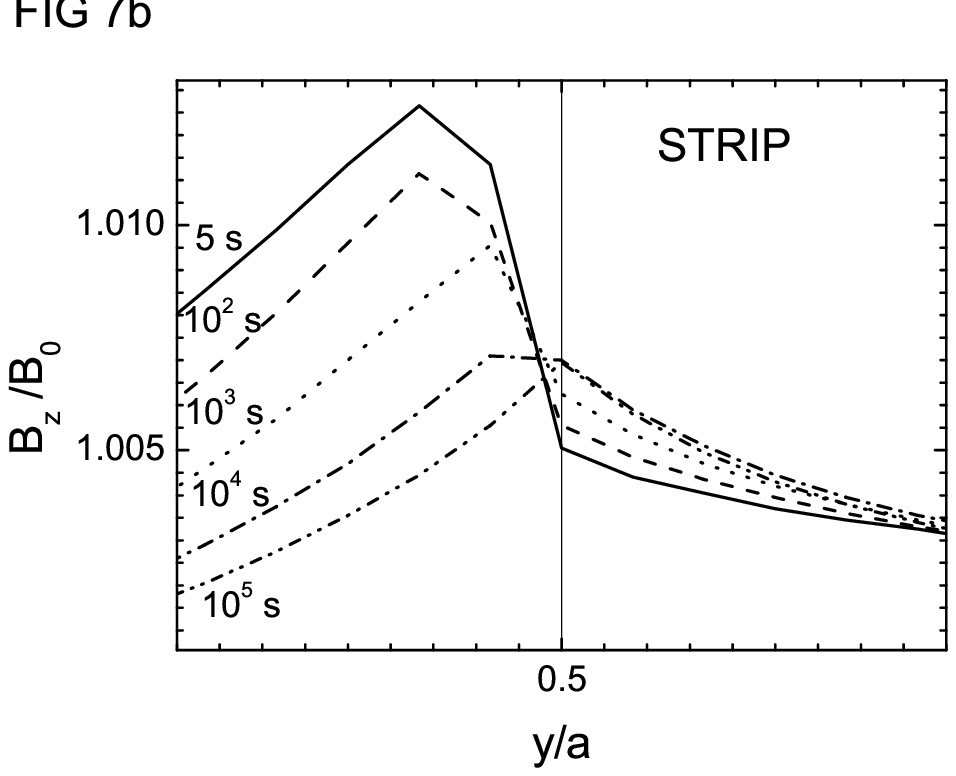}
\caption{zoom of the previous figures, in the region close to the
strip edge.}\label{zoom}
\end{figure}
\begin{figure}[htb]
\includegraphics[width=\figwidth, trim = 0cm 0cm 0cm 0cm, clip]{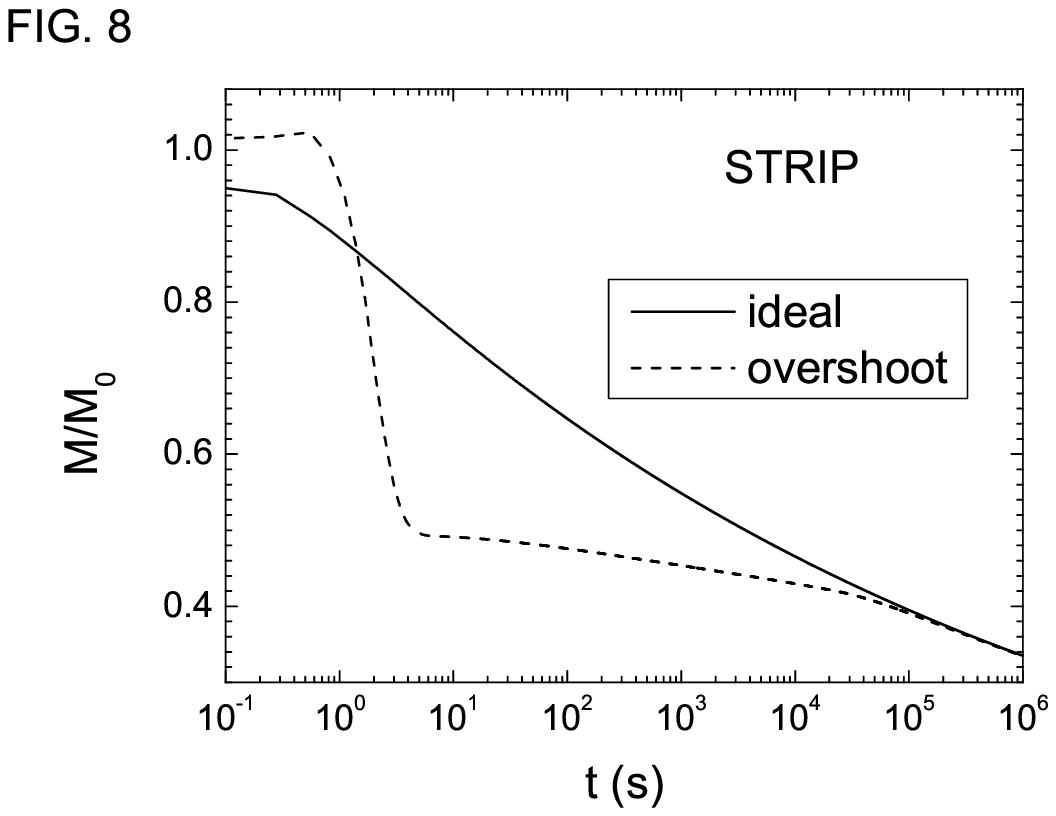}
\caption{Magnetic relaxation curves computed from different
magnetic field ramps in a thick strip.} \label{calcmtnstrip}
\end{figure}
\begin{figure}
\includegraphics[width=\figwidthb, clip]{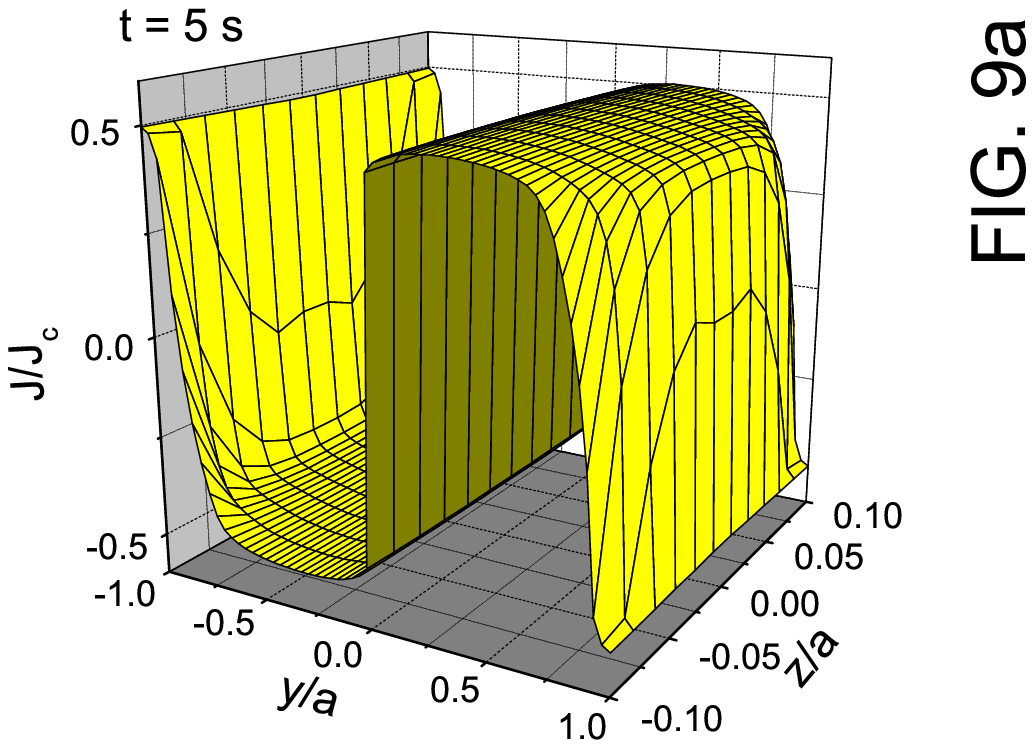}
\includegraphics[width=\figwidthb, clip]{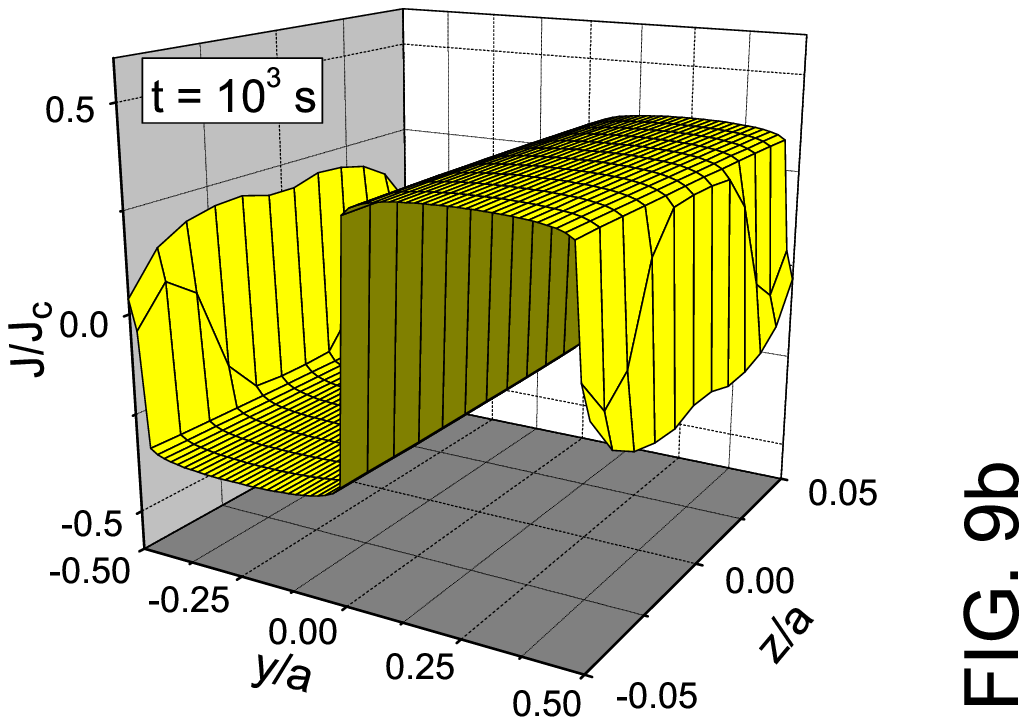}
\includegraphics[width=\figwidthb, clip]{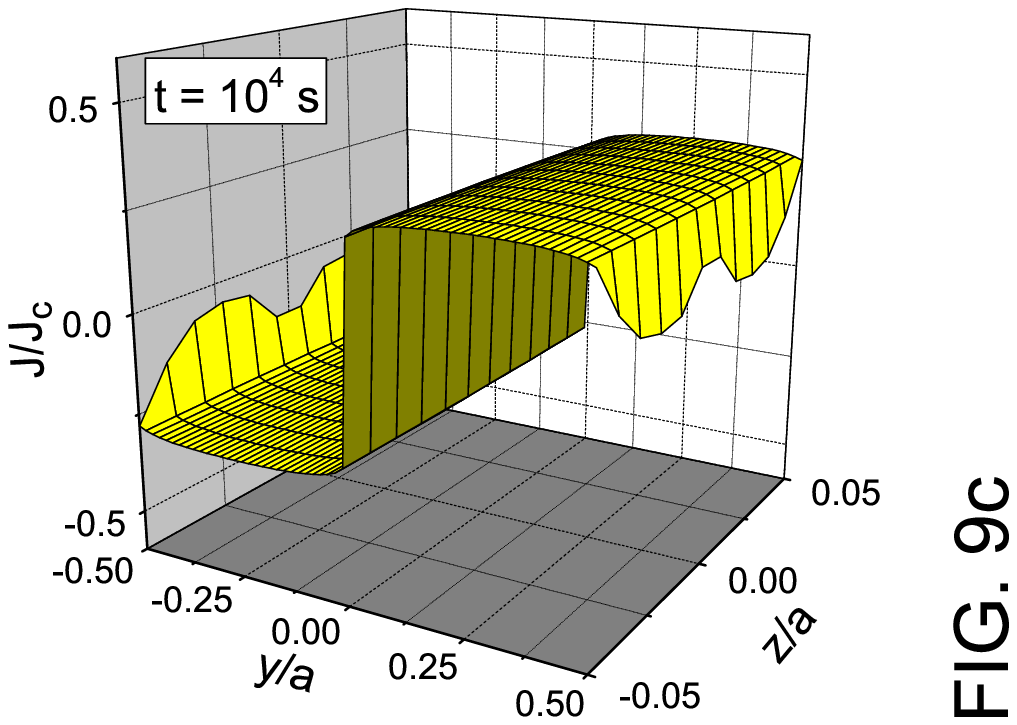}
\includegraphics[width=\figwidthb, clip]{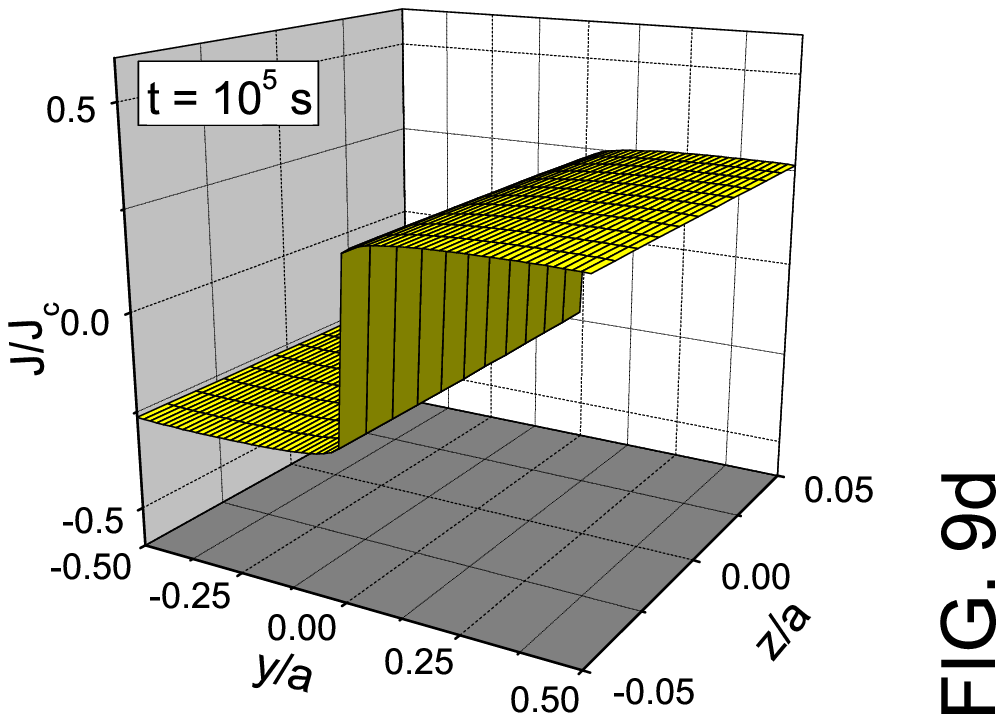}
\caption{Current density distribution at different times for a
strip. The current density is computed in the case of an overshoot
in the field ramp.} \label{jcnt}
\end{figure}
In order to analyze the effect of the sample geometry on the
relaxation, we considered the case of a strip in perpendicular
field for which the main effect of the overshoot  arises on the
surface, where the demagnetizing field is more intense. In Fig.
\ref{profilistripovz0} the magnetic field profiles for a thick
strip ($2a=$1~mm, $2b=$0.1~mm) are reported; a critical current
density of $10^8$~A/m$^2$ and an $E-J$ exponent $n = 15$ are set.
In the upper part of the figure we can see the field profile
relaxations in the ideal case. We can observe that the
demagnetizing field relaxes towards lower magnetic fields on the
surface. At the same time, the field increases in the inner region
and there is a boundary, known as the neutral line, where the
field remains constant; it divides the region with entrapped flux
from the one with shielded flux. If an overshoot of a 1~mT occurs
the flux, as expected, is strongly reduced on the strip edge and
the field maximum is located inside the strip. In the next 10$^6$
seconds the maximum relaxes and moves towards the strip edge
where, at the same time, the field increases. On the contrary, in
the ideal case the field on the border always decreases during the
relaxation. When the maximum reaches the edge, the field profile
in the strip fully resembles the profile computed in the ideal
case and the relaxation continues as in the ideal case. Also in
this case, as shown in the magnetization curves in Fig.
\ref{calcmtnstrip}, the $M(t)$ with and without overshoot join
together at long times. Also in the perpendicular geometry the
evolution of the magnetic state is directed to rebuild a shielded
state. In Fig. \ref{jcnt}, the time evolution of the current
density is shown . During the relaxation the current changes sign
and after  long time the current distribution in the cross section
of the strip rebuilds the distribution of a full shielded state.
Except for the time evolution of the magnetic field on the border
of the strip, in the perpendicular geometry there are not
substantial differences respect to the parallel geometry. In fact,
our computations have shown that in the perpendicular geometry,
for $H_a>H_p$, the demagnetizing effects do not affect the time
evolution of the magnetic relaxation.

\section{\label{sec4}Experimental Results and Discussion}
\begin{figure}[htb]
\includegraphics[width=\figwidth, trim = 0cm 0cm 0cm 0cm, clip]{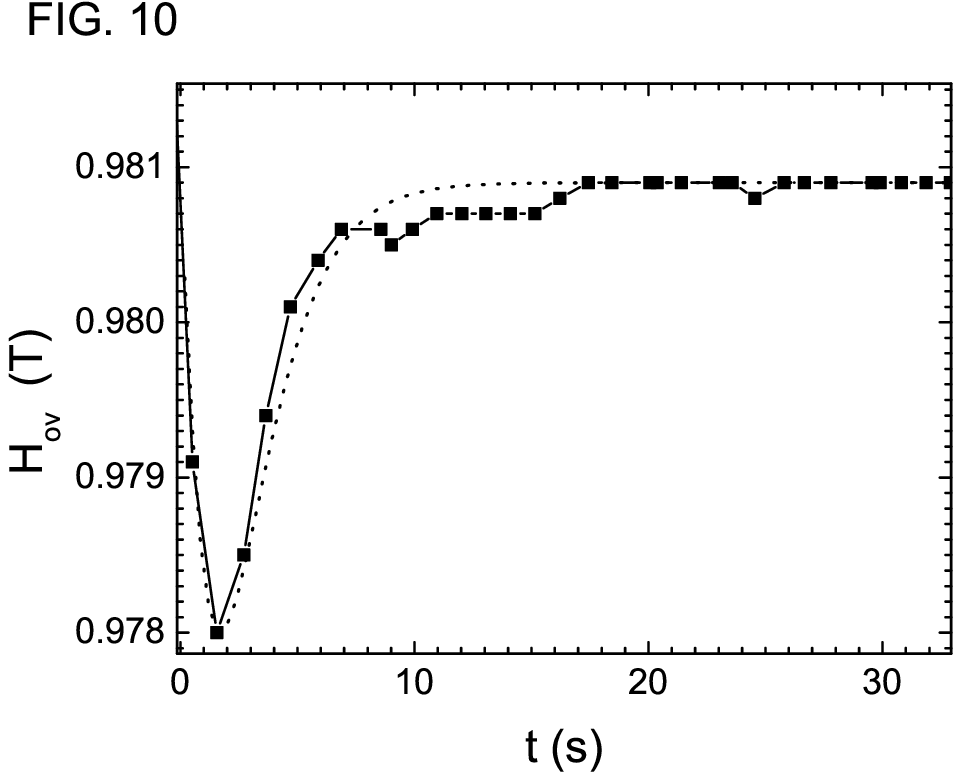}
\caption{Magnetic field ramp with a sweep rate of 3.3 mT/s in the
time window where a field overshoot occurs, as measured by a
hall-probe (square and line), and $H_a(t)$ employed in our
computation (dotted line).} \label{hallprobe}
\end{figure}
\indent Magnetic relaxation measurements have been performed by
means of a Vibrating Sample Magnetometer (VSM) equipped with a 16
T magnet. The external magnetic field can be ramped with a maximum
sweep rate of 7~mT/s. When the field is nominally stopped the
magnetic field has an overshoot of around 1$\div$5~mT depending on
the sweep rate used for ramping the field and this unwanted
feature has been used to induce a ESF state in our samples. We
used a hall probe to measure the time dependence of the external
field and in the inset of the Fig. \ref{hallprobe}  the
measured overshoot for our magnet is shown.\\
\indent In order to check of validity of our numerical results, we
have measured the magnetic relaxation on monofilamentary
BSCCO(2223)/Ag tapes prepared by the standard PIT technique. We
have chosen this kind of  samples because they allow us to study
bulk rectangular samples  with full penetration fields which can
be in the order of 10~mT even at the lowest temperature i.e. 4.2
K. The dimensions of the superconducting region in the measured
sample are 3.02 $\times$ 0.14 $\times$ 4.6~mm$^3$ and the
estimated critical current density ranges from $10^7$ to
$10^9$~A/m$^2$, depending on the
temperature. In this way we can study experimentally the overshoot effects as $H_p$ decreases.\\
\indent $M(t)$ measurements have been performed with  the field
perpendicular to the sample surface ($H \parallel c$-axis) in the
4.2~-~45~K temperature range, cooling the sample in zero-field
(ZFC) for each temperature. The initial magnetic state is obtained
by  increasing $H_a$ with a sweep rate of 3.3~mT/s, up to 2~T.
After this, the field is decreased with the same sweep rate down
to a measuring field $\mu_0H_0 = 1$~T. The field variation of 1~T
is chosen to be, for any measuring temperature, well above $H_p$,
which is evaluated by taking the value of the field corresponding
to the maximum (in absolute value) in the virgin magnetization
curves at 4.2 K. In this way, in absence of a field overshoot, a
full critical state, with entrapped flux, is realized in the
superconductors \cite{RMP68(1996)911}. As the final field $H_0$ is
nominally achieved, the $M(t)$ data are acquired each second for
5000 seconds. \\
\begin{figure}
\includegraphics[width=\figwidth, trim = 0cm 0cm 0cm 0cm,clip]{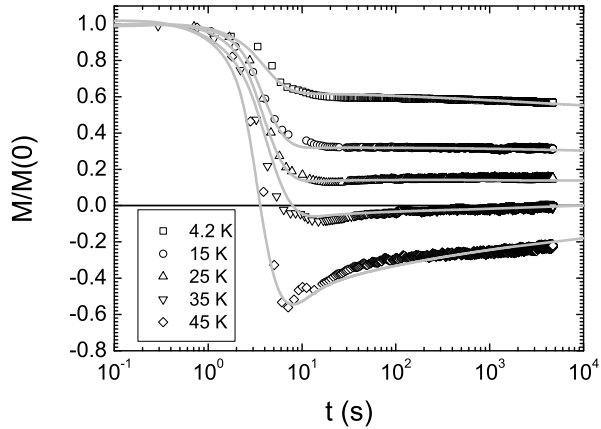}
\caption{Magnetic relaxations measured at different temperatures
for a magnetic field $H_0 =$ 1 T (points) and computed curves
(continuous lines).} \label{mtall}
\end{figure}
\begin{table}
\caption{\label{tab1} critical current densities and exponent $n$
used for the fit of the experimental curves.}
\begin{ruledtabular}
\begin{tabular}{lll}
  $T$ (K) & $J_c$ (A/m$^2$) & $\ n$ \\
  4.2  & $2.40 \ 10^8$ & 20 \\
  15 & $1.15 \ 10^8$ & 19 \\
  25 & $9.70 \ 10^7$ & 13 \\
  35 & $8.70 \ 10^7$ & \ 9 \\
  45 & $4.00 \ 10^7$ & \ 8 \\
\end{tabular}
\end{ruledtabular}
\end{table}
The $M(t)$, normalized at the initial magnetization value $M(0)$,
measured at different temperatures, are shown in Fig. \ref{mtall}.
In all the curves, a large drop in the magnetization occurs during
the first 11 seconds and this time corresponds to time interval
during which the external field has an overshoot. The behaviour of
the magnetization in the subsequent 5000 seconds depends on the
value of the temperature. At 4.2~K the magnetization decreases
slightly, but the relaxation after 5000 seconds does not exhibit
the behaviour expected for a fully entrapped state. At 15~K and
25~K,  the magnetization remains nearly constant, whereas at 35~K
and 45 K the magnetization first takes negative values and then
increases on time. The effect of the overshoot increases as the
full penetration field decreases with the temperature. These
measurements show that the magnetic relaxation can be still
affected by the field overshoot after at list 100 s. The negative
values measured in the $M(t)$ at 45~K mean that the shielded flux
region in the sample is larger than the entrapped
one, although the initial condition was a fully entrapped state.\\
\begin{figure}
\includegraphics[width=\figwidth,  clip]{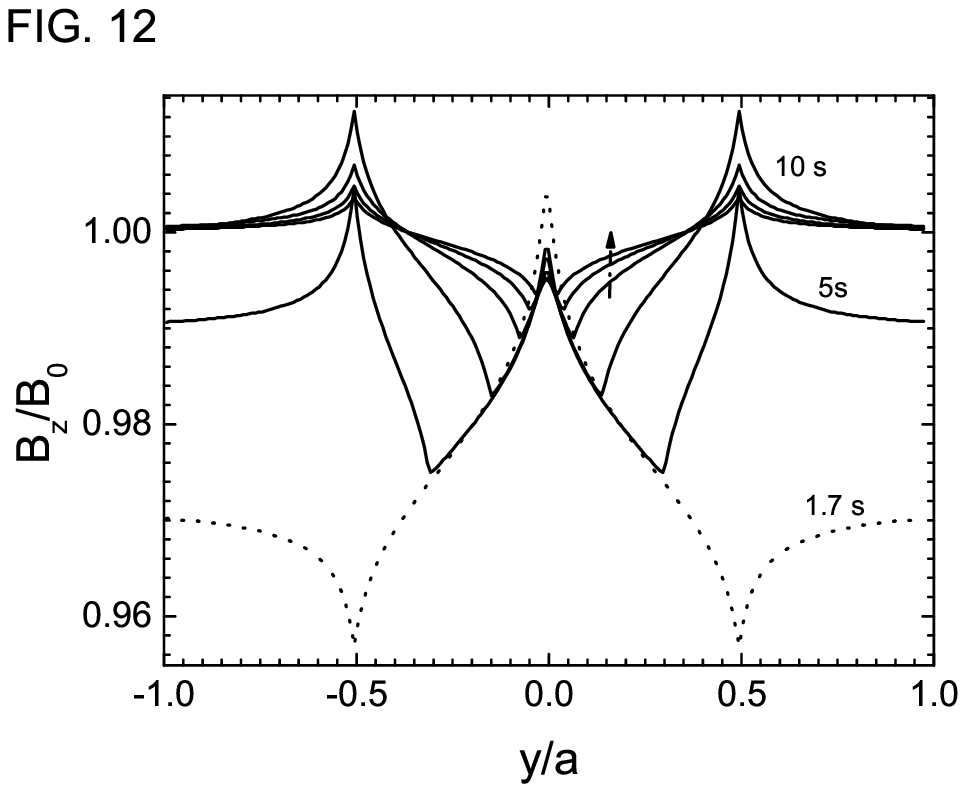}
\caption{Field profiles computed for the relaxation at 45 K. The
profile in the direction of the arrow are computed at $t=$100,
1000, 5000 s.}\label{profili35k}
\end{figure}
\indent In order to reproduce our experimental results, we have
computed the magnetic relaxation for a superconducting strip with
the cross section of our sample. In the computations, the field
ramp reproduces the experimental field ramp, with a sweep rate of
0.0033 T/s. The overshoot has been simulated by using the
exponential function discussed in the Section \ref{sec3}. As shown
in Fig. \ref{hallprobe}, this function reproduces quite well the
experimental overshoot with $H_{ov} = 0.029$~mT, $t_{ovm} = 1.7$~s
and $c = 1.3$. In our computation we have to set both $n$ and
$J_c$. The exponent $n$ has been evaluated by measuring the
hysteresis loop at different sweep rate. Taking the $M$ values
measured at 1~T for different sweep rate ($\dot{B_a}$), $n$ is
given by means of a linear fit of $\log(\dot{B_a})$ as function of
$\log(M)$; the $n$ values reported in Tab. \ref{tab1} have been
rounded to the nearest integer. On the other hand, the critical
current density is a free parameter chosen in order to obtain the
best fit. From our computations, it results $J_c = 2.4 \ 10^8$
A/m$^2$ at 4.2 K and  $4.0 \ 10^7$~A/m$^2$ at 45 K. As shown in
Fig. \ref{mtall}, the numerical computations  reproduce well the
experimental behaviour. In Fig. \ref{profili35k}, the profile
computed at T=45 K are shown. In particular, at t=10 s, when $H_a$
is practically constant, it results that the magnetic state in the
superconductor has both the regions with entrapped and shielded
flux. In the next 5000 seconds, the profile relaxes toward a
shielded state, which is practically fulfilled at t=5000 s, when
the simulation is stopped.\\
\indent Our work shows that the first seconds of the relaxation
have to be analyzed very carefully in order to estimate correctly
the creep rate and, thus, extract information about the pinning
properties of the sample. In fact, our results show that it is not
appropriate just to cut the first seconds of the relaxation curves
and extract information from the remanent data if the presence of
an overshoot in the magnet has not been previously considered.

\section{\label{sec6}Conclusion}
In this work, we have studied the magnetic relaxation from a state
with shielded and entrapped flux, generated by a field overshoot
after the nominal stop of the external field. The magnetic
relaxations have been computed in parallel and perpendicular
geometry. The computed magnetization shows a large drop in the
first seconds due to the flux expulsion from the samples boundary.
After long time, the $M(t)$ curves computed with and without field
overshoot (having, thus, as initial condition an ESF and a full
shielded or entrapped flux state, respectively) join together.
Moreover, our simulations show that, during the relaxation, the
same value of the magnetization corresponds to the same magnetic
state. In addition, the experimental relaxation curves, measured
on BSCCO(2223) tapes, are well reproduced by our numerical
computations, allowing us to correctly analyze the $M(t)$ from the
instant when the external field is nominally stopped.

\begin{acknowledgments}
We thank A. Ferrentino and G. Perna for their technical support.
\end{acknowledgments}

\bibliography{esfstates}

\end{document}